\documentclass{article}

\usepackage{amsmath,mathrsfs}
\usepackage{ upgreek }
\usepackage{ amssymb }
\usepackage{graphicx,subfigure}
\graphicspath{{v-phi/}}
\usepackage{multirow,booktabs}
\usepackage[numbers,sort&compress]{natbib}

\def\abstract#1{\vskip 7mm 
        \begin{center}{\large Abstract}\par \smallskip
                \begin{minipage}[c]{12cm}
                        \small #1
                \end{minipage}
        \end{center}
}
\def\title#1{\begin{center}{\Large\bf #1}\end{center}}
\def\author#1{\vskip 5mm \begin{center}{#1}\end{center}}
\def\address#1{\begin{center}{\it #1}\end{center}}
\makeatletter
\@ifundefined{lesssim}{}{}
\@ifundefined{gtrsim}{}{}
\def\vereq#1#2{\lower3pt\vbox{\baselineskip1.5pt \lineskip1.5pt
\ialign{$\m@th#1\hfill##\hfil$\crcr#2\crcr\sim\crcr}}}
\makeatother

\begin{document}

\title{	The Effective Potential Originating from Swampland and the Non-trivial Brans-Dicke Coupling
}
\author{ Qi Li$^a$, Jing Li$^a$, Yongxiang Zhou$^a$ and Xun Xue$^{a,b,}\footnote{Corresponding author:xxue@phy.ecnu.edu.cn}$}
\address{ $^a$Department of Physics, East China Normal University, Shanghai 200241, China}
\address{ $^b$Center for Theoretical Physics, Xinjiang University, Urumqi 830046, China}
\abstract{
	The effective vacuum energy density contributed by the non-trivial contortion distribution and the bare vacuum energy density can be viewed as the energy density of the auxiliary quintessence field potential. We find that the negative bare vacuum energy density from string landscape leads to a monotonically decreasing quintessence potential while the positive one from swampland leads to the meta stable or stable de Sitter like potential. Moreover, the non-trivial Brans-Dicke like coupling between quintessence field and gravitation field is necessary in the latter case.
	}

\section{Introduction}
\begin{figure}[htbp]
	\centering
	\subfigure[]
	{\label{ads}
		\includegraphics[width=3in]{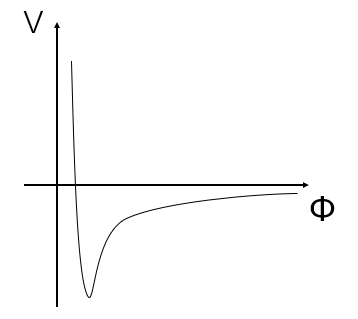}}
	\subfigure[]
	{\label{ds}
		\includegraphics[width=3in]{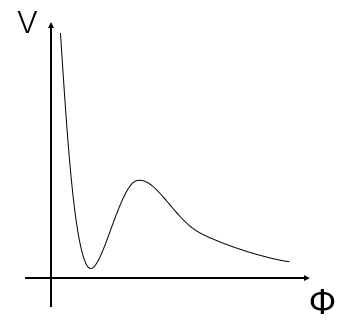}}
	\caption{illustrations of (a) the AdS vacuum and (b) the metastable dS vacuum}\label{addss}
\end{figure}

String theory is believed to be one of the candidates for the theories of quantum gravity. The reduction from string theory to the low energy field theory relies on the choice of possible ways of string vacua compactification, which can be of order $10^{500}$ and makes the string theory lack of prediction power. The string vacua that preserve supersymmetry are generally with negative energy density and hence are the anti-de Sitter(AdS) ones. However, one needs to lift the vacuum energy to positive (see the illustration Figures \ref{ads} and \ref{ds}) to explain the turning up of the inflation and the late time accelerating expansion of the universe, e.g. by the KKLT construction\cite{Kachru:2003aw}. The second criterion of swampland conjecture excludes the effective field theory(EFT) with a meta-stable dS vacuum as theory possessing ultraviolet(UV) completion\cite{Obied:2018sgi,Brennan:2017rbf} and leaves the quintessence type of potential as the only possible one to account for the evolution of the universe. 

Recently, an alternative construction to lift the AdS vacua to de Sitter(dS) type is proposed by utilizing the frozen large scale Lorentz violation which results in the contortion distribution at the cosmic scale\cite{Zhai:2019aps}. It is pointed out that the possible Lorentz violation in quantum gravity during the era dominated by quantum gravity before inflation may be frozen by inflation. The scale of Lorentz violating area may be stretched outside of horizon\cite{wu2015effective,Shen:2018elj} and it may be transformed into a cosmic one even when the scale is reenter the horizon during the late time expansion. The dark energy may be an emergent effect of the large scale Lorentz violated effective gravity at the cosmic scale \cite{Shen:2018elj} in contrary to many proposed candidates for dark energy ranging from cosmological constant model to the dynamical dark energy model such as quintessence model\cite{Tsujikawa:2013fta} etc.

\begin{figure}[htbp]
	\centering
		\includegraphics[width=3in]{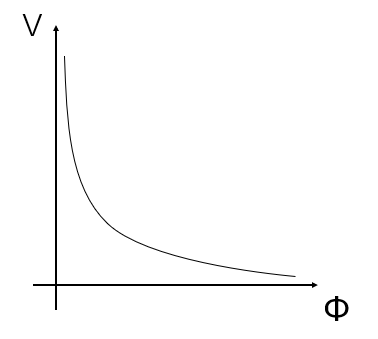}
	\caption{The monotonically decreasing quintessence potential can meet the requirement of the swampland conjecture and the accelerating expansion of the universe.}\label{unds}
\end{figure}

In quintessence model, it is supposed that the spacetime of the universe is dominated by an auxiliary scalar, namely quintessence, field with the potential decreasing in magnitude as the value of the scalar field increasing, as indicated in Figure \ref{unds}. Although quintessence model is successful in describing the late time accelerating expansion of the universe, its origin is somewhat ambiguous. Is it fundamental or an effective description of some other quantities which are more original within the framework of fundamental physics? In this paper, the effective vacuum energy contributed by bare cosmological constant and contortion defined in\cite{Zhai:2019aps} is described by the quintessence field potential energy effectively. 

The evolution of the effective vacuum energy density proposed in the work by Zhai et. al.\cite{Zhai:2019aps} shows a transition from monotonically decreasing behaviour when the bare vacuum energy density comes from string landscape to the existence of a local minimum with one from swampland. The result seems to be  consistent with the requirement by the second criterion in swampland conjecture. However, the equivalent behaviour between evolution of the effective vacuum energy density and the potential of an auxiliary field such as quintessence field is not obvious. 

Rather than the conventional quintessence model, we consider that the quintessence field may interact with gravity in a non-minimal way, e.g. the Brans-Dicke coupling, in general. We repeat the similar basic picture  given in \cite{Zhai:2019aps} that there is a transition for the quintessence potential from monotonically decreasing to the meta-stable or stable de Sitter one which possesses a local minimum when the bare vacuum energy density varies from negative as in the landscape to positive as in swampland. Moreover, we find the non-trivial Brans-Dicke coupling between quintessence field and gravity is necessary in the swampland case.

\section{The Modified Friedmann Equations by Cosmic Contortion }

The effective gravitation theory with large scale Lorentz violation can be realized by the introduction of gauge fixing constrain to the Hilbert-Einstein theory. The constrain terms in the action supply an effective spin angular momentum distribution to the equations of motion for connection even in the case of scalar matter source. The connection is thus with torsion in general and deviated from Levi-Civita one by the contortion part. The contortion part contributes in turn an effective energy-momentum distribution to the equations of motion for the metric tensor, the modified Einstein equations,  with the Einstein tensor composed purely by Levi-Civita connection at the left hand side of the equations and the organizing of the terms involving contortion and bare cosmological constant to the right hand side of the equations\cite{wu2015effective,wu2016sim,yang2017VSR,wei2017E2,Shen:2018elj}. Taking the possible cosmic scale Lorentz boost violation into account, the modification to Einstein equations is the dark partner energy-momentum contribution from contortion in addition to the matter source one\cite{Shen:2018elj},

\begin{equation}
\widetilde{R}_{\ c}^{a}-\frac{1}{2}\delta_{\ c}^{a}\widetilde{R}=8\pi G\left( T+T_{\Lambda}\right)_{\ c}^{a} 
\end{equation}
where $\widetilde{R}_{\ c}^{a}$ is the Ricci tensor in Levi-Civita connection and $T_{\Lambda}$ is the dark partner contribution of energy-momentum tensor, $[T_{\Lambda}]_{\ c}^{a}=diag(\rho_{\Lambda}, -p_{\Lambda}, -p_{\Lambda}, -p_{\Lambda})$. The energy density $\rho_{\Lambda}$ and pressure $p_{\Lambda}$ can be expressed as 
\begin{equation}
\rho_{\Lambda}=-\dfrac{c^4}{8\pi G}(3\mathscr{K}^2+6\mathscr{K}\dfrac{\dot{a}}{a}-\Lambda_0)
\end{equation}
and 
\begin{equation}
 p_{\Lambda}=-\dfrac{c^4}{8\pi G}(\mathscr{K}^2+4\mathscr{K}\dfrac{\dot{a}}{a}+2\dot{\mathscr{K}}-\Lambda_0)
\end{equation}
respectively, where $\mathscr K(t)={K^0}_{11}={K^0}_{22}={K^0}_{33}$, the only non-zero components of contortion in a Robertson-Walker solution ansatz for the metric tensor of a uniform and isotropic universe in the comoving frame, and $\Lambda_0$ is the bare cosmological constant from vacuum energy density. Denote $\Lambda$ as the cosmological constant fitted from observation with $\Lambda CDM$ model and $x=\dfrac{\Lambda_{0}}{\Lambda}$. The modified Friedmann Equation in geometrical unit $\dfrac{c^4}{8G\pi}=1$ can be written as\cite{Shen:2018elj},

\begin{equation}
\mathscr{K}^2 + 2\mathscr{K}\dfrac{\dot{a}}{a} +\left( \dfrac{\dot{a}}{a}\right) ^2=\dfrac{1}{3}\left( \rho+\Lambda_{0}\right) 
\label{0mFriedmann}
\end{equation}
and 
\begin{equation}
\ddot{a}=-\frac{a}{2}\left( p+\dfrac{\rho}{3}\right) +\dfrac{1}{3}a\Lambda_{0}-\dfrac{d}{dt}\left( a\mathscr{K}\right) 
\label{imFriedmann}
\end{equation}
or
\begin{equation}
\dot{H}(t)+\dot{\mathscr{K}}(t)+H(t)\left[ H(t)+\mathscr{K}(t)\right]+\dfrac{3w+1}{2}\left[ H(t)+\mathscr{K}(t)\right] ^2-\dfrac{w+1}{2}\Lambda_{0}=0
\end{equation}
with the utilizing of equation of state for cosmic media $p=w\rho$.

The evolution of $\mathscr{K}(t)$ is fixed by three kinds of approximation utilizing one independent equation of Friedmann equations in $\Lambda CDM$ model or the equation of state for dark partner part, to close the set of equations in addition to the modified Friedmann equations.  
The three approximations are Case A with additional equation
\begin{equation}
\dot{\mathscr{K}}=\dfrac 13\Lambda_0-\dfrac 13\Lambda-H\mathscr{K}\;\;,
\end{equation}
Case B with
\begin{equation}
\dot{\mathscr{K}}+(3w+2)H\mathscr{K}+\dfrac {3w+1}2\mathscr{K}^2=\frac {w+1}2(\Lambda_0-\Lambda)
\end{equation}
and Case C with
\begin{equation}
(3w_0+1)\mathscr{K}^2+(6w_0+4)H\mathscr{K}+2\dot{\mathscr{K}}=(w_0+1)\Lambda_0
\end{equation}
where $p_{\Lambda}=w_0\rho_{\Lambda}$, respectively\cite{Shen:2018elj}.

The initial value of $\mathscr{K}(t)$ is obtained in a similar way,
\begin{equation}
\mathscr{K}(t_0)=H_0\left( \pm\sqrt{1-\dfrac{\Lambda-\Lambda_{0}}{3H_0}}-1\right) 
\end{equation}
and hence a constrain for the value of $\Lambda_0$
\begin{equation}
\Lambda_{0}\geqslant-(3H_0-\Lambda)\approx-\dfrac{2}{5}\Lambda 
\end{equation}
The evolution of effective cosmological constant or the energy density of the dark partner, 
\begin{equation}
\Lambda_{eff}=\Lambda_0-3(\mathscr{K}^2+2\mathscr{K}\dfrac{\dot{a}}{a})\;\;,
\end{equation}
exhibits that there is a critical value for $\Lambda_0$ which separates the monotonically decreasing $\Lambda_{eff}$ phase from one with a local minimum. The critical value $\Lambda_{0crit} \approx 0$ is just the division of vacuum energy density from landscape and swampland\cite{Zhai:2019std}. 

\begin{figure}[h]
	\centering
	\subfigure[]
	{\label{lambda_eff_caseA1}
		\includegraphics[width=2.5in]{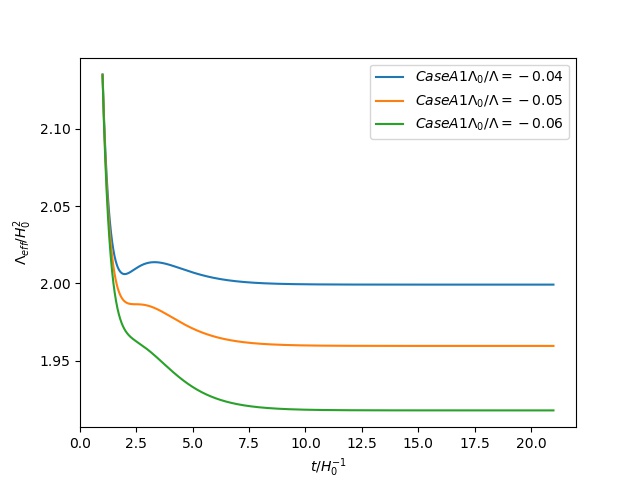}}
	\subfigure[]
	{\label{lambda_eff_caseB1}
		\includegraphics[width=2.5in]{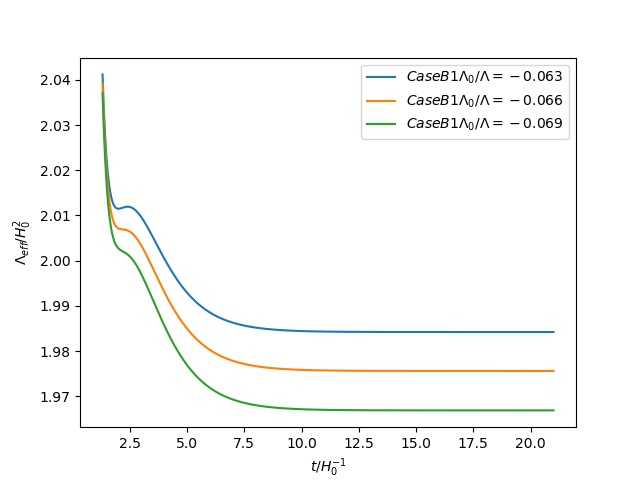}}
	{\label{lambda_eff_caseC1}
		\includegraphics[width=2.5in]{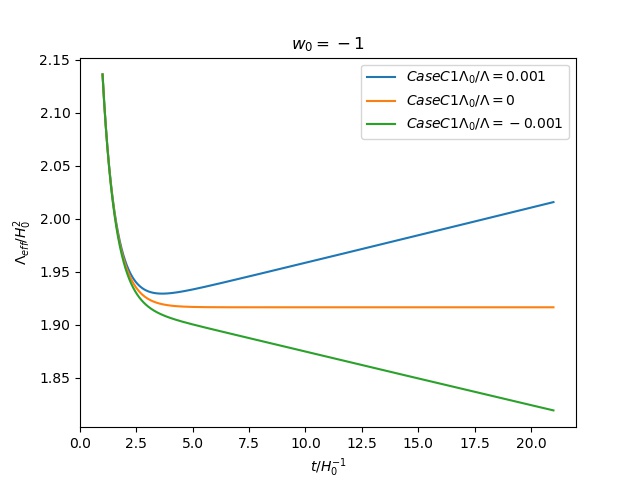}}
	\subfigure[]
	{\label{lambda_eff_caseC1w=89}
		\includegraphics[width=2.5in]{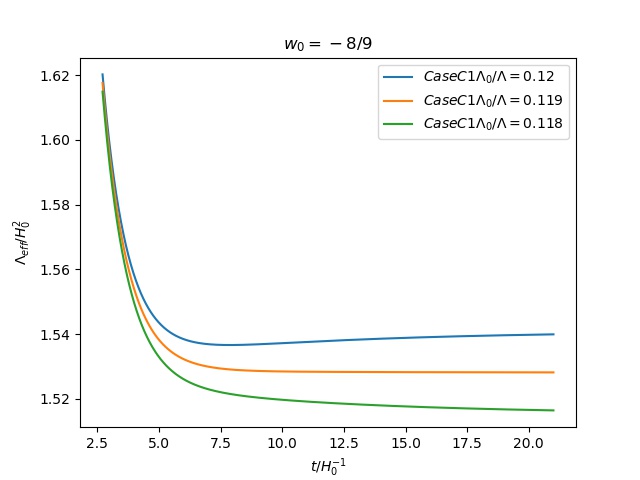}}
	\caption{The transition of $\Lambda_{eff}$ from a monotonically decreasing quintessence phase to a phase with local minimum in case of  $\mathscr{K}(t_0)=H_0\left( \sqrt{1-\dfrac{\Lambda-\Lambda_{0}}{3H_0}}-1\right) $ in \cite{Zhai:2019std}.}\label{lambdaeff}
\end{figure}

\begin{figure}[h]
	\centering
	\subfigure[]
	{\label{lambda_eff_caseA2}
		\includegraphics[width=2.5in]{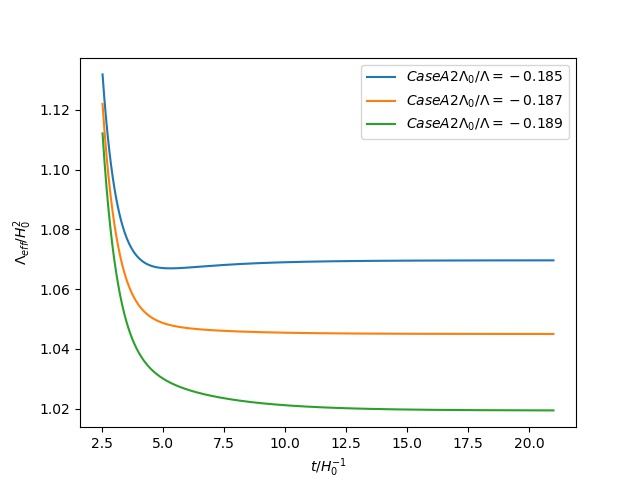}}
	\subfigure[]
	{\label{lambda_eff_caseB2}
		\includegraphics[width=2.5in]{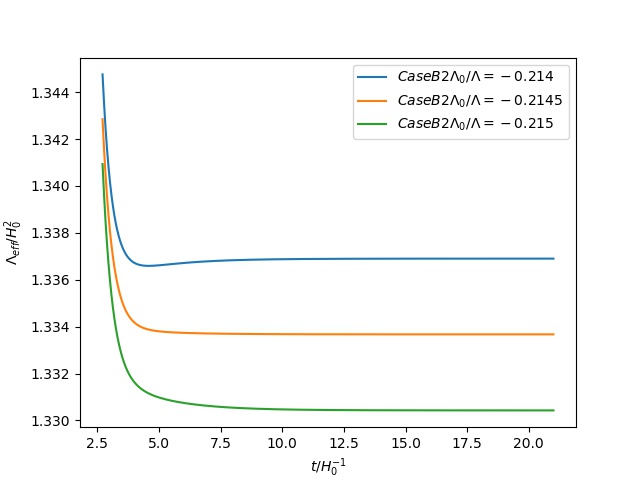}}
	\subfigure[]
	{\label{lambda_eff_caseC2}
		\includegraphics[width=2.5in]{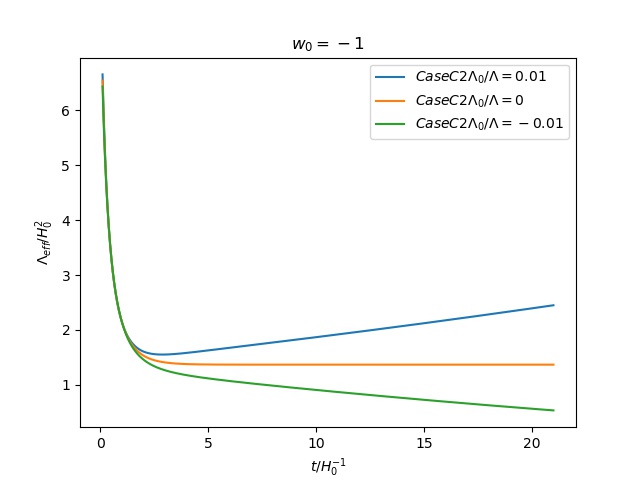}}
	\subfigure[]
	{\label{lambda_eff_caseC2w=89}
		\includegraphics[width=2.5in]{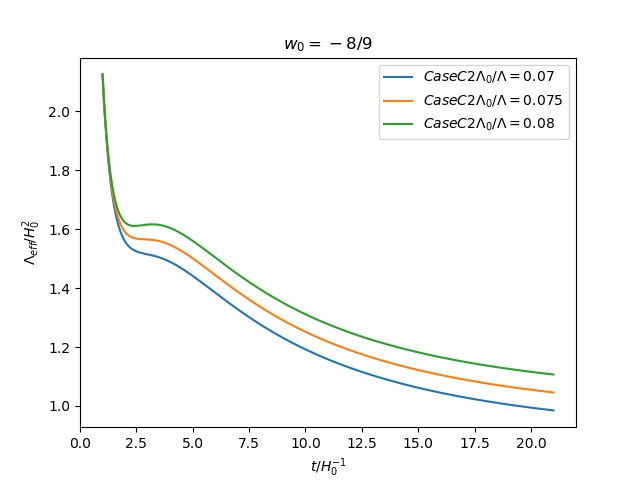}}
	\caption{The transition of $\Lambda_{eff}$ from a monotonically decreasing quintessence phase to a phase with local minimum in case of  $\mathscr{K}(t_0)=H_0\left( -\sqrt{1-\dfrac{\Lambda-\Lambda_{0}}{3H_0}}-1\right) $ in \cite{Zhai:2019std}.}\label{lambdaeff2}
\end{figure}

\section{The Effective Potential and the Non-trivial Brans-Dicke Coupling}
\begin{figure}[h]
	\centering
	\subfigure[]
	{\label{caseAxi_k+}
		\includegraphics[width=2.8in]{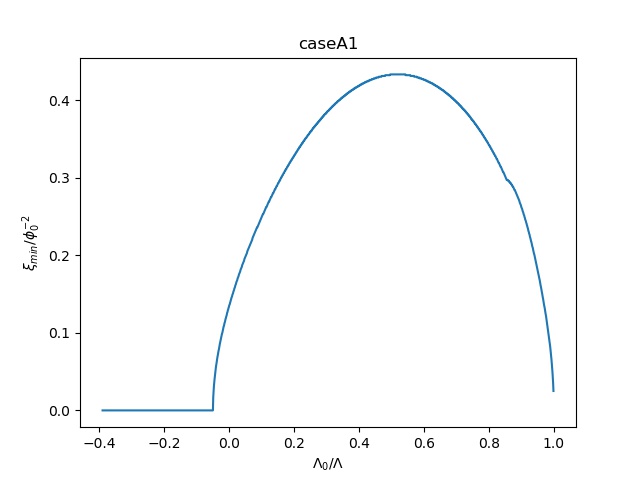}
	}
	\subfigure[]
	{\label{caseBxi_k+}
		\includegraphics[width=2.8in]{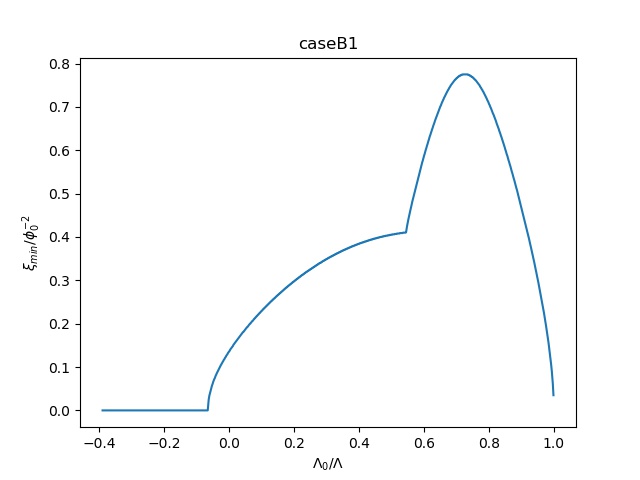}
	}
	\subfigure[]
	{\label{caseCxi_k+}
		\includegraphics[width=2.8in]{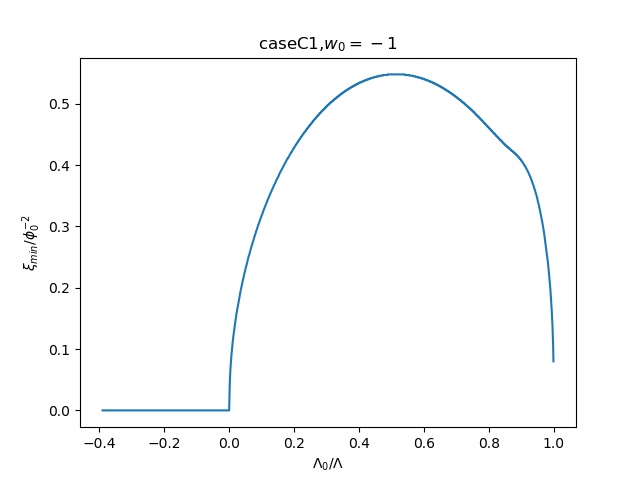}
	}
	\subfigure[]
	{\label{caseCxi_k+98}
		\includegraphics[width=2.8in]{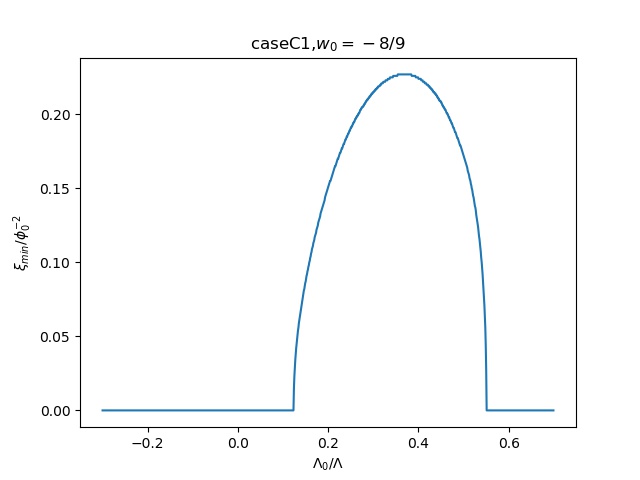}
	}
	\caption{ The cases of initial value  $\mathscr{K}(t_0)=\sqrt{1-(\Lambda-\Lambda_0)/3H_0^2}-H_0$ revealing  the non-triviality of the non-minimal coupling, $\xi_{min}>0$, when $\Lambda_0>\Lambda_{0-crit}$}\label{xicasec1}
\end{figure}

\begin{figure}[h]
	\centering
	\subfigure[]
	{\label{caseAxi_k-}
		\includegraphics[width=2.8in]{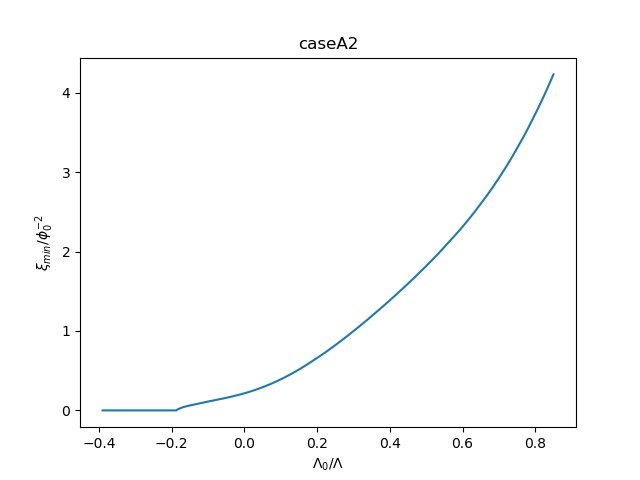}
	}
	\subfigure[]
	{\label{caseBxi_k-}
		\includegraphics[width=2.8in]{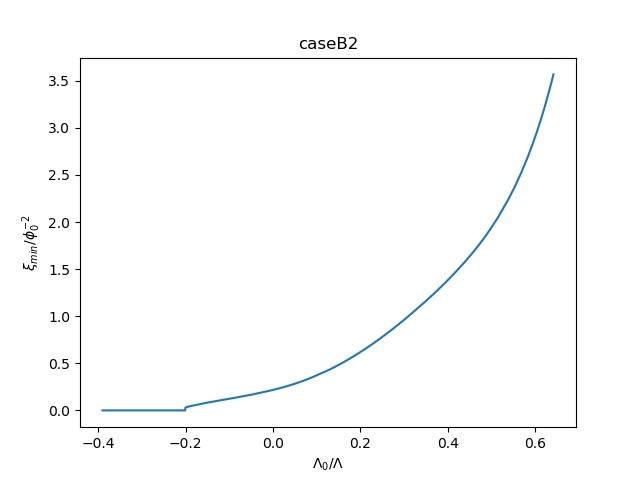}
	}
	\subfigure[]
	{\label{caseCxi_k-}
		\includegraphics[width=2.8in]{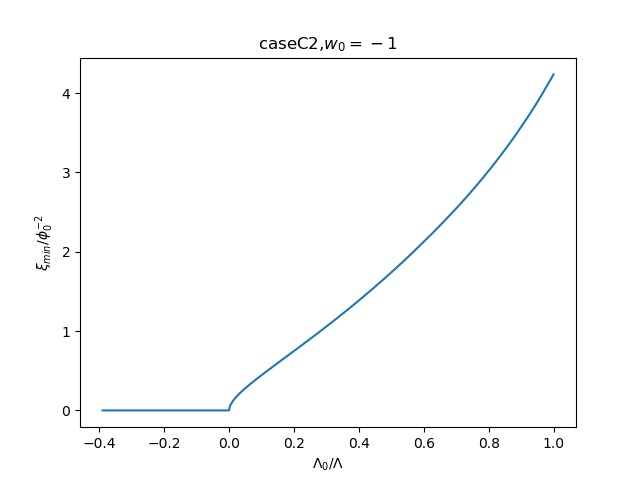}
	}
	\subfigure[]
	{\label{caseCxi_k-98}
		\includegraphics[width=2.8in]{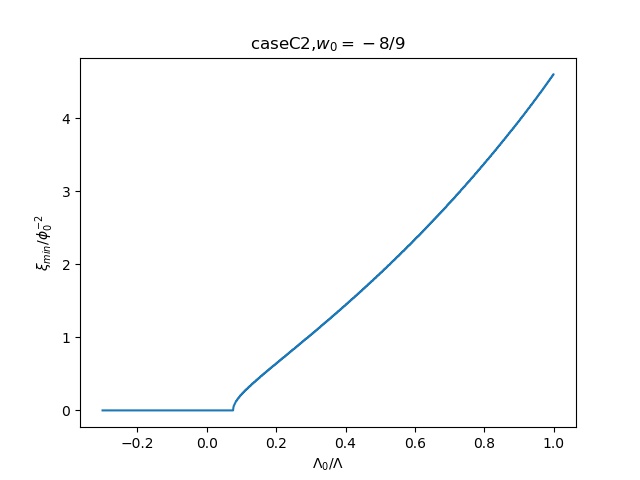}
	}
	\caption{ The cases of initial value $\mathscr{K}(t_0)=-\sqrt{1-(\Lambda-\Lambda_0)/3H_0^2}-H_0$ revealing  the non-triviality of the non-minimal coupling, $\xi_{min}>0$, when $\Lambda_0>\Lambda_{0-crit}$}\label{xicasec2}
\end{figure}
In quintessence model, the dark energy energy-momentum tensor is described by the one contributed by quintessence field\cite{Tsujikawa:2013fta}. If the quintessence field is regarded as an effective description of the contortion effect caused by large scale Lorentz violation, $\Lambda_{eff}$, the energy density of the dark partner, can be represented by the energy density produced by quintessence field. Consider a model of gravity involving non-minimal Brans-Dicke type of coupling between gravity and quintessence field,

\begin{equation}
S=\int d^4x\sqrt{-g}\left[\frac 12M^2_{pl}R\left( 1+\xi\phi^2\right) -\frac 12g^{\mu\nu}\partial_{\mu}\phi\partial_{\nu}\phi-V(\phi)\right]+S_m
\label{noncoupling}
\end{equation}
where $M_{pl}$ is the reduced Planck mass, $R$ is the Ricci scalar, $\xi$ is the non-minimal coupling constant between gravity and quintessence field and $S_m$ is the matter action. The equations of motion for gravity and scalar field are
\begin{equation}
\left( 1+\xi\phi^2\right) \left( R_{\mu\nu}-\frac{1}{2}g_{\mu\nu}R\right) =8\pi GT_{\mu\nu}
\end{equation}
and
\begin{equation}
V_{,\phi}+3H\dot{\phi}+\ddot{\phi}-M_{pl}^2R\xi\phi=0
\label{mequ}
\end{equation}
respectively. 
The pressure and energy density of quintessence are given, by $P_{\phi}=\left[ \dot{\phi}^2/2-V(\phi)\right] /(1+\xi\phi^2)$ and $\rho_{\phi}=\left[ \dot{\phi}^2/2+V(\phi)\right] /(1+\xi\phi^2)$, respectively. In an effective description of the dark partner energy density by quintessence field, we have

\begin{equation}
\Lambda_{eff}=\rho_{\phi}=\frac {\dot{\phi}^2+2V(\phi)}{2+2\xi\phi^2}
\label{26}
\end{equation}
or
\begin{equation}
\dot{\Lambda}_{eff}\left( 1+\xi\phi^2\right)+2\Lambda_{eff}\xi\phi\dot{\phi}+3H\dot{\phi}^2-M_{pl}^2R\xi\phi\dot{\phi}=0
\end{equation}
by Eq.(\ref{mequ}), where the metric is assigned with Robertson-Walker metric in the form of $ds^2=-dt^2+a^2(t)\left(dr^2+r^2d\Omega^2\right)$.
Formally, we have 
\begin{equation}
\dot{\phi}=M_{pl}^2\xi\phi\left( \dfrac{\dot{H}}{H}+2H\right)-\dfrac{\Lambda_{eff}\xi\phi}{3H}\pm\sqrt{\left[ M_{pl}^2\xi\phi\left( \dfrac{\dot{H}}{H}+2H\right)-\dfrac{\Lambda_{eff}\xi\phi}{3H}\right]^2-\dfrac{\dot{\Lambda}_{eff}}{3H}\left( 1+\xi\phi^2\right)  } 
\label{l_xi}
\end{equation}
which gives a constrain on the Brans-Dicke coupling $\xi$ as
\begin{equation}
\left[3M_{pl}^2\left(\dot{H}+2H^2\right)-\Lambda_{eff}\right]^2\phi^2\xi^2-3H\dot{\Lambda}_{eff}\phi^2\xi-3H\dot{\Lambda}_{eff}\geqslant 0
\label{36}
\end{equation}

\begin{figure}[h]
	\centering
	\subfigure[]
	{\label{phi-t-caseA_k+}
	\includegraphics[width=2.8in]{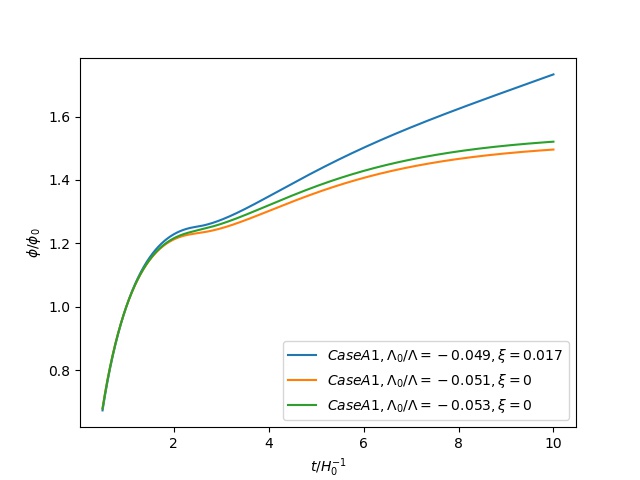}}
	\subfigure[]
	{\label{phi-t-caseB_k+}
	\includegraphics[width=2.8in]{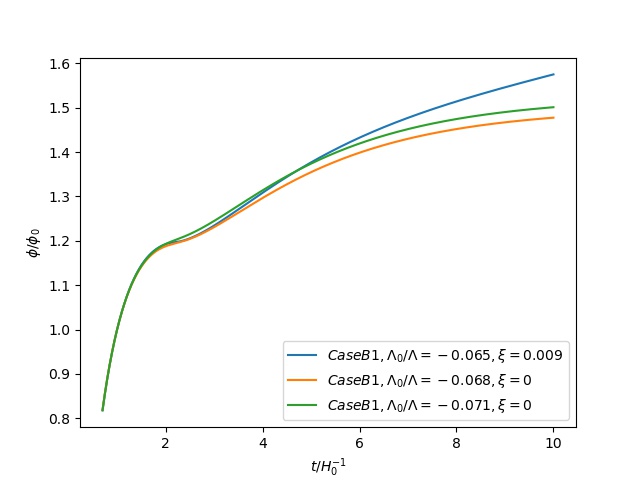}}
	\subfigure[]
	{\label{phi-t-caseC_k+}
	\includegraphics[width=2.8in]{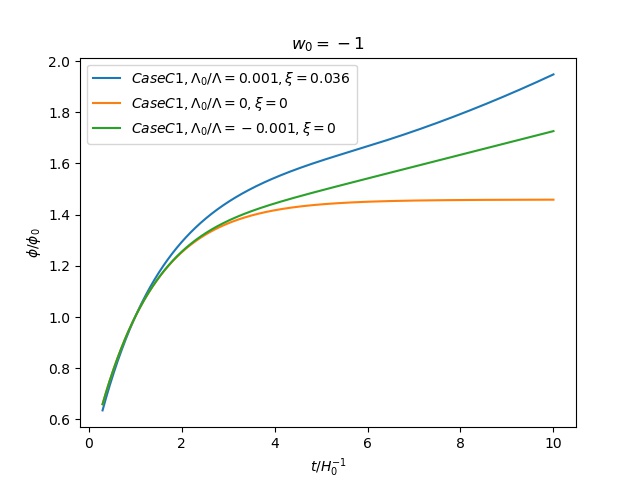}}
	\subfigure[]
	{\label{phi-t-caseC_k+89}
	\includegraphics[width=2.8in]{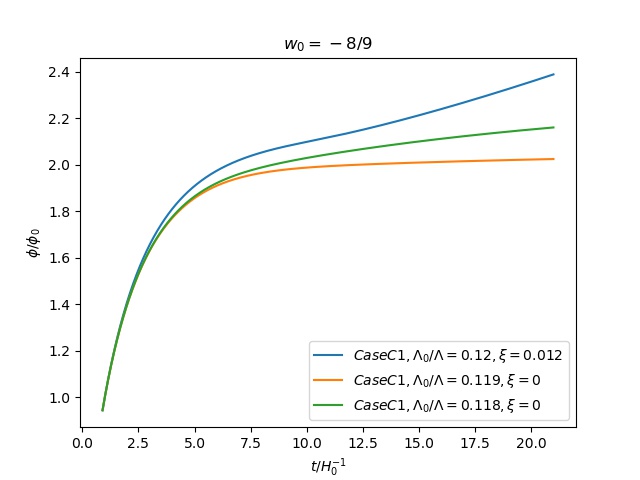}}
	\caption{The monotonic evolution of $\phi(t)$ in cases with initial value $\mathscr{K}(t_0)=H_0( \sqrt{1-(\Lambda-\Lambda_{0})/3H_0}-1) $ }\label{phit1}
\end{figure}

\begin{figure}[h]
	\centering
	\subfigure[]
	{\label{phi-t-caseA_k-}
		\includegraphics[width=2.8in]{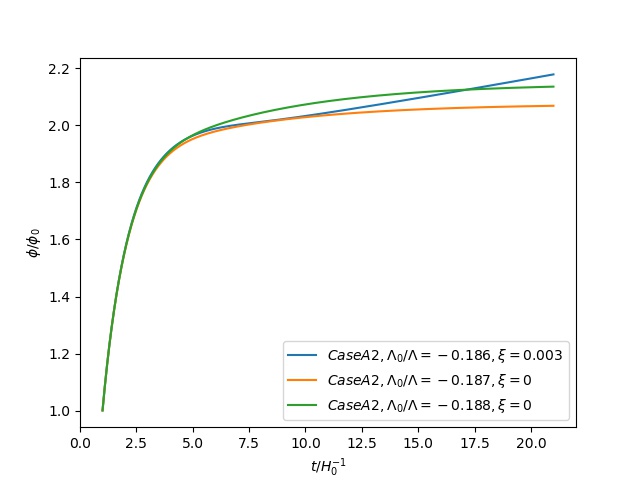}}
	\subfigure[]
	{\label{phi-t-caseB_k-}
		\includegraphics[width=2.8in]{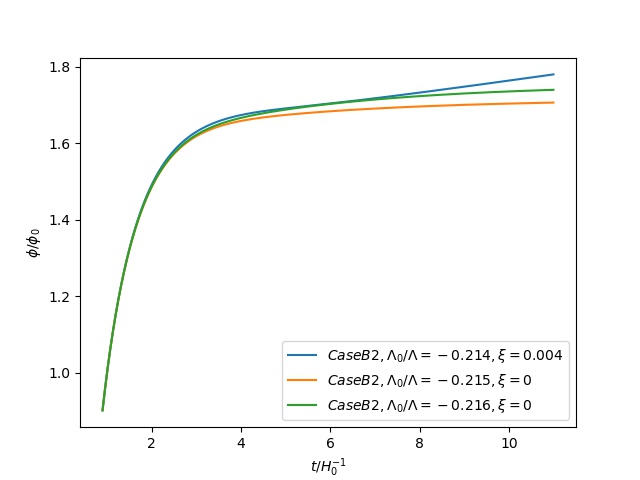}}
	\subfigure[]
	{\label{phi-t-caseC_k-}
		\includegraphics[width=2.8in]{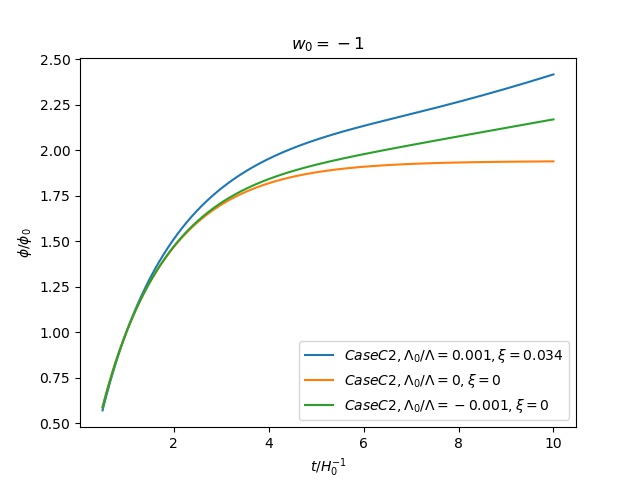}}
	\subfigure[]
	{\label{phi-t-caseC_k-89}
		\includegraphics[width=2.8in]{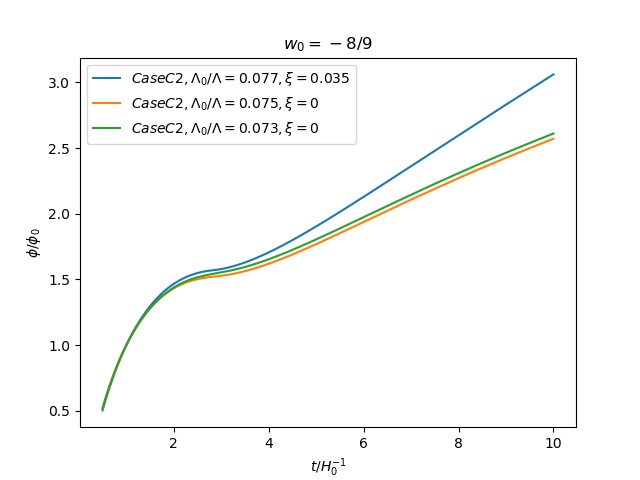}}
	\caption{The monotonic evolution of $\phi(t)$ in cases  with  $\mathscr{K}(t_0)=H_0( -\sqrt{1-(\Lambda-\Lambda_{0})/3H_0}-1) $ }\label{phit2}
\end{figure}

\begin{figure}[htbp]
	\centering
	\subfigure[]
	{\label{caseA_k+}
		\includegraphics[width=2.8in]{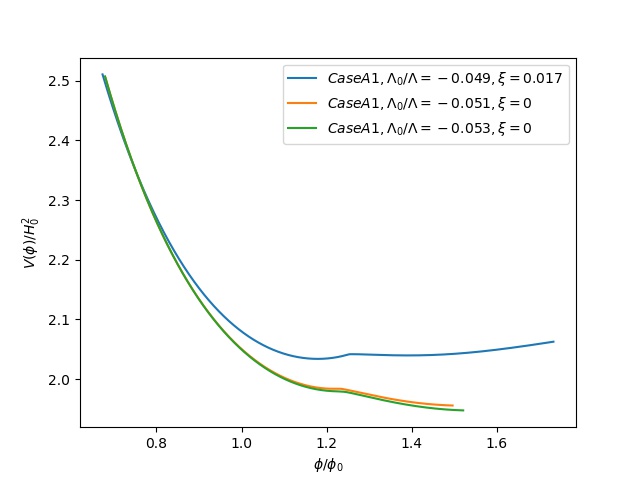}}
	\subfigure[]
	{\label{caseB_k+}
		\includegraphics[width=2.8in]{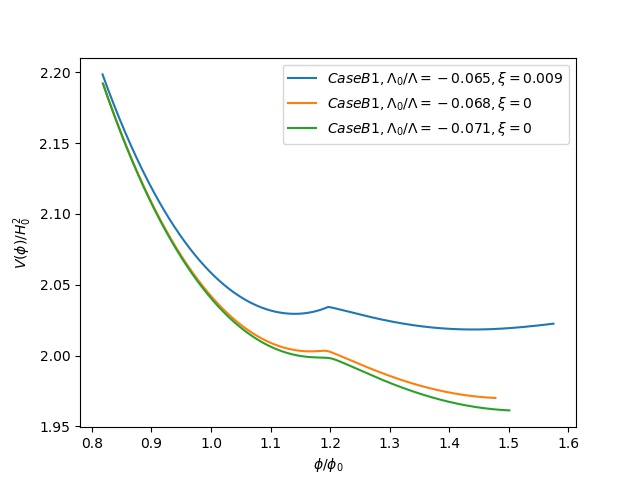}}
	\subfigure[]
	{\label{caseC_k+}
		\includegraphics[width=2.8in]{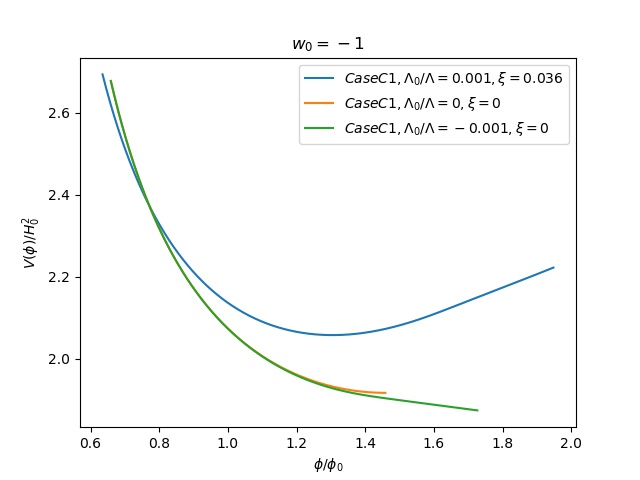}}
	\subfigure[]
	{\label{caseC_k+89}
		\includegraphics[width=2.8in]{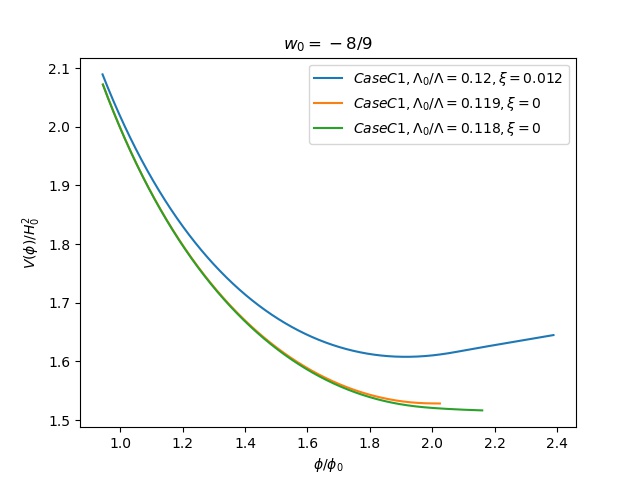}}
	\caption{$V(\phi)$ shifts from a monotonically decreasing quintessence phase to a dS one with a local minimum versus $\phi$ evolution in cases of $\mathscr{K}(t_0)=H_0( \sqrt{1-(\Lambda-\Lambda_{0})/3H_0}-1) $}\label{caseA1}
\end{figure}

\begin{figure}[htbp]
	\centering
	\subfigure[]
	{\label{caseA_k-}
		\includegraphics[width=2.8in]{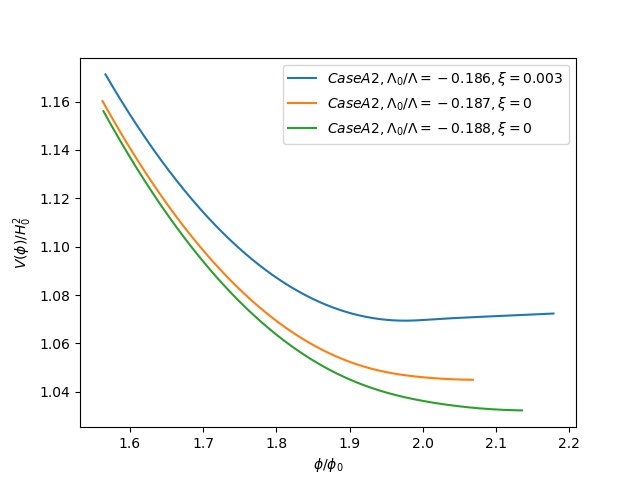}}
	\subfigure[]
	{\label{caseB_k-}
		\includegraphics[width=2.8in]{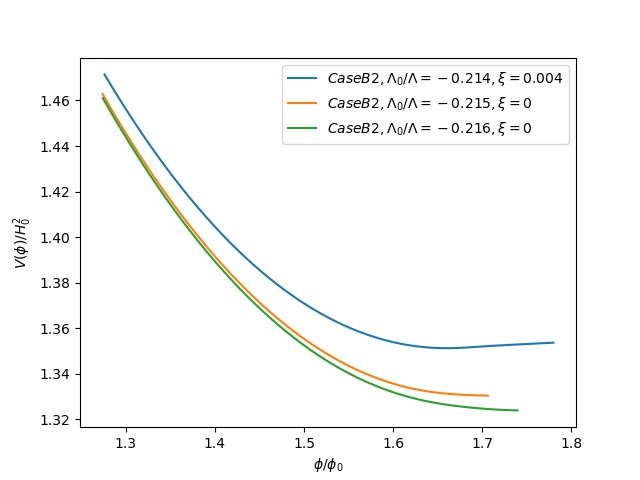}}
	\subfigure[]
	{\label{caseC_k-}
		\includegraphics[width=2.8in]{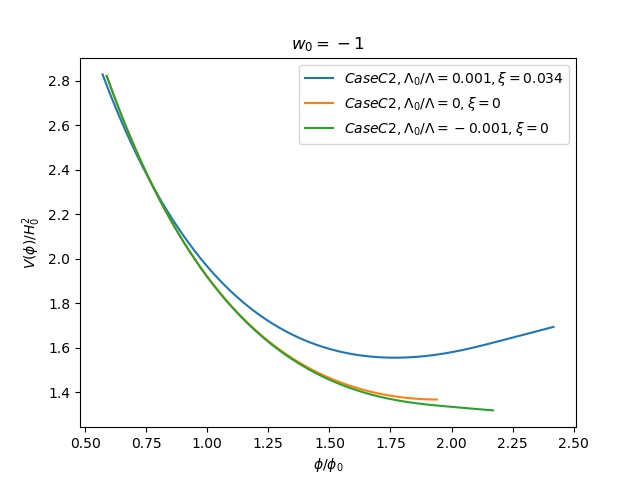}}
	\subfigure[]
	{\label{caseC_k-89}
		\includegraphics[width=2.8in]{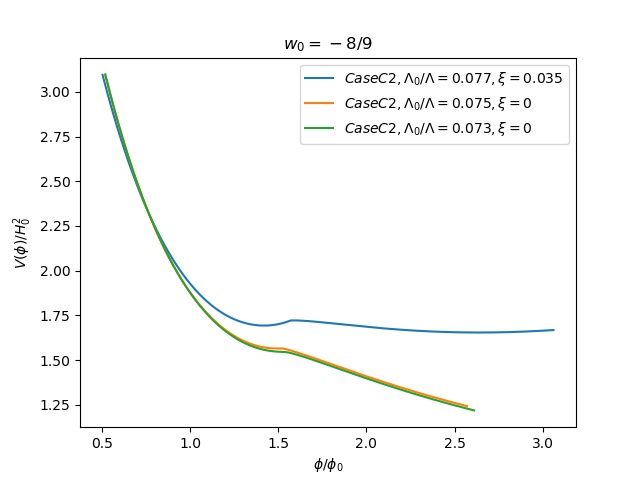}}
	\caption{$V(\phi)$ shifts from a monotonically decreasing quintessence phase to a dS one with a local minimum versus $\phi$ evolution in cases of $\mathscr{K}(t_0)=H_0( -\sqrt{1-(\Lambda-\Lambda_{0})/3H_0}-1) $}\label{caseA2}
\end{figure}

The minimal values of the Brans-Dicke coupling constant $\xi_{min}$ in all cases are plotted in Figure \ref{xicasec1} and Figure \ref{xicasec2}. It reveals that the non-trivial Brans-Dicke type of coupling is necessary when $\Lambda_0$ exceeds a critical value in every case.

The numerical calculation of Eq.(\ref{l_xi}) with plus sign gives the evolution of $\phi$, a monotonically evolution versus $t$, as is shown in Figure \ref{phit1} and Figure \ref{phit2}. The field $\phi$ may thus be regarded as an intrinsic time for the evolution of the universe in some sense. The evolution of $V(\phi)$ versus $\phi$ is shown in Figure \ref{caseA1} and Figure \ref{caseA2}. The critical value for $\Lambda_{0}$ which symbolizes the transformation of $V(\phi)$ from a monotonically quintessence like to the meta-stable dS potential is almost the same as one for $\Lambda_{eff}(t)$ in \cite{Zhai:2019std} in most cases.

It is a universal conclusion that the minimal Brans-Dicke coupling is non-trivial, i.e. $\xi_{min}>0$, in all the cases when $\Lambda_0>\Lambda_{crit}$. Noting that the minimal value of the coupling constant satisfies
\begin{equation}
\dot{\Lambda}_{eff}=\dfrac{\dot{\phi}\left( \ddot{\phi}+V_{,\phi}-2\xi\phi\Lambda_{eff}\right) }{1+\xi\phi^2}=\dfrac{\dot{\phi}\left[ \left( M^2_{pl}R-2\Lambda_{eff}\right)\xi\phi-3H\dot{\phi} \right] }{1+\xi\phi^2}
\end{equation}
from Eq.(\ref{36}), in the increasing region of $V(\phi)$ when $\Lambda_0>\Lambda_{crit}$, $\dot{\Lambda}_{eff}>0$ implies that 
\begin{equation}
 \left( M^2_{pl}R-2\Lambda_{eff}\right)\xi\phi-3H\dot{\phi} \geqslant 0
\end{equation}
At the minimum of $V(\phi)$, it holds that
\begin{equation}
M^2_{pl}R-2\Lambda_{eff} > 0
\end{equation}
i.e.
\begin{equation}
\xi\geqslant \dfrac{3H\dot{\phi}}{\left(M^2_{pl}R-2\Lambda_{eff}\right)\phi}
\end{equation}


As is pointed out in \cite{Zhai:2019std}, the Case C approximation is not a good one from the comparison between Hubble constant versus time and the luminosity distance versus redshift $z$. 
The reason may be that a fixed $w_0$ in the equation of state of dark partner part is assumed. The case $w_0>-8/9$ can be ignored (excluded by observation of luminosity distance with redshift relation\cite{Zhai:2019std}), we can make the conclusion that
the meta-stable dS potential needs non-trivial coupling between quintessence field and gravitation. 
The conclusion that quintessence potential can be generated from string landscape AdS vacuum effectively and the critical value of cosmological constant separating quintessence from meta-stable dS is approximately zero still holds for the effective description of quintessence field theory.

\section{Summary and Outlook}

Our conclusion that for string landscape with $\Lambda_0>-(3H_0-\Lambda)\approx-2/5\Lambda$, the effective cosmological constant naturally give a quintessence like potential is consistent with second criterion of the dS Swampland conjecture\cite{Obied:2018sgi,Brennan:2017rbf}. The uplifting of AdS to a positive effective cosmological constant by frozen large scale Lorentz violation mechanism is equivalent to the uplifting of the negative energy density of the AdS vacuum to a quintessential evolving energy density in the quintessence field description. The approach avoids the meta-stable dS swampland puzzle and has a quantum gravity origin. For string swampland with positive cosmological constant for most reasonably approximation, the effective quintessence field potential behaves as a metastable dS potential and the non-trivial Brans-Dicke type of coupling is necessary while the minimal Brans-Dicke coupling constant can vanish in the landscape case. 

The non-minimal coupling between gravity and scalar field originates from the Machian idea of variational gravitational constant which is developed into Brans-Dicke theory as a competitor of general relativity\cite{PhysRev.124.925, PhysRev.125.2194}. The motivation of including a non-minimally coupling of inflaton to gravity in inflationary cosmology is that the Brans-Dicke type of coupling can support the assumption of identification of inflaton with the stand model Higgs particle\cite{PhysRevD.40.1753, PhysRevD.41.1783, Bezrukov:2007ep}. Despite of phenomenological impact on the observational constraints in cosmological application, the non-minimal coupling is of theoretical necessity for several theoretical motivations. Actually, the quantum correction to a scalar field theory coupling with gravity would generate such a term automatically and it is a consistency requirement to include the term in the renormalization procedure. From an effective field theory point of view, such a term should be treated with equal footing with the Hilbert-Einstein term for it is marginal in a renormalization group treatment of the theory\cite{Steinwachs:2019hdr}. In the inflation model, the term can induce an asymptotic scale invariance for the large field value which provides inflation in a natural way. Moreover, scalar fields such as the dilaton or the moduli field in the string inspired effective theories would inevitably interact with gravity in a non-minimally way. Our results show that the non-minimal coupling in the effective description by the quintessence field theory is only necessary in the swampland phase during the late time accelerating expansion while it supplies the naturally description of the turning on and off for inflation. Though the appearance of non-minimal coupling is universal from the inflationary era to the late time evolution, it falls to different phases for inflaton field from quintessence field. The non-minimal coupling may be absent in the phase originating from landscape. The effective quintessence scalar field description of large scale distribution of contortion has two different phases as our conclusion leads to, one with vanishing minimal Brans-Dicke coupling constant corresponding to the landscape vacua and the other with non-trivial Brans-Dicke coupling constant corresponding to the swampland vacua. As the Brans-Dicke coupling term is marginal in the renormalization group scheme, it hints us that it may be worthy to investigate whether the effective scalar field theory description of late time accelerating expansion can reveal different fixed point solutions in the renormalization group approach or not.

\section*{Acknowledgment}
This work is supported by the National Natural Science Foundation of China, under Grant No. 11775080 and Grant No. 11865016.


\bibliographystyle{utcaps}
\bibliography{References}

\end{document}